\documentclass[submission,copyright,creativecommons]{eptcs}
\providecommand{\event}{ } 
\usepackage{breakurl}             
\usepackage{amssymb,stmaryrd,amsmath}
\usepackage{pgf}

\usepackage{color}

\providecommand{\urlalt}[2]{\href{#1}{#2}}
\providecommand{\doi}[1]{doi:\urlalt{http://dx.doi.org/#1}{#1}}

\newcommand{\mathrmL}{{\mathchoice{\mbox{\rm\L}}{\mbox{\rm\L}}{\mbox{\rm\scriptsize\L}}{\mbox{\rm\tiny\L}}}}

\newcommand{\sfP}{\mathsf{P}}
\newcommand{\sfV}{\mathsf{V}}
\newcommand{\sfS}{\mathsf{S}}
\newcommand{\sfM}{\mathsf{M}}
\newcommand{\sfE}{\mathsf{E}}

\newcommand{\lpii}{\rm \L\Pi\frac{1}{2}}

\newcommand{\cp}{\odot}
\newcommand{\tp}{\to_{\Pi}}
\newcommand{\chalf}{\overline{\frac{1}{2}}}

\newcommand{\f}{\ensuremath{\varphi}}
\newcommand{\p}{\ensuremath{\psi}}
\newcommand{\x}{\ensuremath{\chi}}

\newcommand{\conj}{\ensuremath{\mathbin{\&}}}
\newcommand{\lpi}{{\rm \mathrmL\Pi}}
\newcommand{\half}{{\frac{1}{2}}}

\newtheorem{theorem}{Theorem}[section]

\newtheorem{proposition}[theorem]{Proposition}
\newtheorem{definition}[theorem]{Definition}

\title{Games for the Strategic Influence of Expectations}
\author{Llu\'is Godo
\institute{Artificial Intelligence Research Institute, IIIA\\ 
Spanish National Research Council, CSIC\\
Campus UAB, 08193 Bellaterra, Spain}
\email{godo@iiia.csic.es}
\and
Enrico Marchioni
\institute{Institut de Recherche en Informatique de Toulouse\\
Universit\'e Paul Sabatier\\
118 Route de Narbonne, 31062 Toulouse, France}
\email{enrico.marchioni@irit.fr}
}
\def\titlerunning{Games for the Strategic Influence of Expectations}
\def\authorrunning{L. Godo, E. Marchioni}
\begin{document}
\maketitle

\begin{abstract}
We introduce a new class of games where each player's aim is to randomise her strategic choices in order to affect 
the other players' expectations aside from her own. The way each player intends to exert this influence is 
expressed through a Boolean combination of polynomial equalities and inequalities with rational coefficients.
We offer a logical representation of these games as well as a computational study of the existence of equilibria.\footnote{This extended abstract is based on the article \cite{GoMa13} and an upcoming extended
version of the same work.}
\end{abstract}

\section{Introduction}

In the situations of strategic interactions modelled in Game Theory, 
the goal of each player is essentially the maximisation of her own expected payoff.
Players, however, often care not only about maximising their own expectation,
but also about influencing other players' expected outcomes.
As an example, consider a number of competing 
investment banks selling and buying tradable assets
so that the trading of financial products affects each other's profit.
These banks might randomize their choices and obviously aim
at maximizing their expected profit. Still, their strategy might go beyond the
choice of a specific investment and they might be interested in influencing the market
and the behavior of other banks possibly undermining the expected gain of their competitors.

In this work, we offer logical models to formalize these kinds of
strategic interactions, called Expectation Games, 
where each player's aim is to randomise her strategic choices in order to affect the other players' expectations over an outcome as well as their own expectation.
Expectation Games are an extension of \L ukasiewicz games \cite{MaWo13} and are based on the logics
$\mathsf{E}({\mathfrak{G}})$ that formalise reasoning about expected payoffs in a class of \L ukasiewicz games \cite{GoMa13}.
{\L}ukasiewicz games \cite{MaWo13},
a generalisation of Boolean games \cite{HaHpMeWi01},
involve a finite set of players $P_i$
each controlling a finite set of propositional variables $\sfV_i$, whose strategy
corresponds to assigning values from the scale
$L_k=\left\{0, \frac{1}{k}, \dots, \frac{k-1}{k}, 1\right\}$ to the variables in $\sfV_i$.
Strategies can be interpreted as efforts or costs, and each player's strategic choice can be seen as
an assignment to each controlled variable carrying an intrinsic cost.
Each player is given a {finitely}-valued \L ukasiewicz {logic} formula $\varphi_i$, with variables from $\bigcup^n_i\sfV_i$,
whose valuation is interpreted as the payoff function for $P_i$ and corresponds to
the restriction over $L_k$ of a continuous piecewise linear polynomial function \cite{CiMuOt99}.

Expectation Games
expand Lukasiewicz games by assigning to each player $P_i$ a modal formula $\Phi_i$ of the logic $\mathsf{E}({\mathfrak{G}})$, whose interpretation corresponds to a piecewise rational polynomial function whose variables are interpreted as the expected values
of the payoff functions $\varphi_i$.
Each formula $\Phi_i$ is then meant to represent a player's goal concerning the relation between her and other players' expectations.

\section{Logical Background}\label{sec:Background}

{  The language of {\L}ukasiewicz logic {\L} (see \cite{CiMuOt99}) is built from a countable set of 
propositional variables $\{p_1,p_2,\ldots\}$, the binary connective $\to$ and the 
truth constant $\overline{0}$ (for falsity).}
Further connectives are defined as follows:
\begin{center} 
\begin{tabular}{rllcrll}
$\neg \varphi$ &is& $\varphi \to \bar{0}$,& &
$\varphi \land \psi$ &is& $\varphi \& (\varphi \to \psi)$,\\
$\varphi \& \psi$ &is& $\neg(\varphi \to\neg \psi)$,  & &
$\varphi \lor \psi$ &is& $((\varphi \to\psi)\to\psi)$,\\
$\varphi \oplus \psi$& is &$ \neg(\neg\varphi \& \neg\psi)$,& &
$\varphi \leftrightarrow \psi$& is &$ (\varphi \to \psi) \& (\psi \to \varphi)$,\\
$\varphi \ominus \psi$ &is& $\varphi\& \neg\psi$,  & &
$d(\varphi, \psi)$& is &$ \neg(\varphi\leftrightarrow \psi)$. \\
\end{tabular}
\end{center}

Let {\em Form} denote the set of {\L}ukasiewicz logic formulas. A valuation $e$ from {\em Form}
into $[0, 1]$ 
is a mapping $e:$ {\em Form} $\to [0, 1]$ assigning to all
propositional variables a value from the real unit interval (with
$e(\overline{0})=0$) that can be extended to
complex formulas as follows:
\[
\begin{array}{r c l c r c l}
e(\varphi \rightarrow \psi)& = & \min(1-e(\varphi)+e(\psi),1) & &
e(\neg \varphi)& = & 1 - e(\varphi) \\
e(\varphi \& \psi)& = &\max(0, e(\varphi)+e(\psi)-1)  & &
e(\varphi \oplus \psi) & = & \min(1, e(\varphi)+e(\psi))  \\
e(\varphi \ominus \psi)& = &\max(0, e(\varphi)-e(\psi))  & &
e(\varphi \land \psi) & = & \min(e(\varphi), e(\psi))  \\
e(\varphi \lor \psi) & = & \max(e(\varphi), e(\psi))  & &
e(d(\varphi, \psi))& = & \vert e(\varphi)- e(\psi)\vert  \\
e(\varphi \leftrightarrow \psi)& = & 1 - \vert e(\varphi)- e(\psi)\vert   & & & & \\
\end{array}
\]

\noindent A valuation $e$ {\em satisfies} a formula $\varphi$ if $e(\varphi)=1$.
As usual, a set of formulas is called a theory.
A valuation $e$ satisfies a theory $T$, if $e(\psi)=1$,
for every $\psi\in T$.

Infinite-valued {\L}ukasiewicz logic has the following axiomatisation:\smallskip

\begin{tabular}{ll}({\L}1) $\varphi\to(\psi\to\varphi)$, & ({\L}2) $(\varphi\to\psi)\to((\psi\to\chi)\to(\varphi\to\chi))$, \\({\L}3) $(\neg\varphi\to\neg\psi)\to (\psi\to\varphi)$, & ({\L}4) $((\varphi\to\psi)\to\psi)\to((\psi\to\varphi)\to\varphi)$.\end{tabular}\smallskip

\noindent The only inference rule is {\em modus ponens}, i.e.: from $\varphi\to\psi$ and
$\varphi$ derive $\psi$.

A {\em proof} in {\L} is a sequence $\varphi_1, \dots, \varphi_n$
of formulas such that each $\varphi_i$ either is an axiom of {\L}
or follows from some preceding $\varphi_j, \varphi_k\ (j, k<i)$ by modus ponens.
We say that a formula 
$\varphi$ can be derived from a theory $T$, denoted as
$T\vdash \varphi$, if there is a proof of $\varphi$ 
from a set
$T'\subseteq T$. A theory $T$ is said to be consistent if $T\not\vdash\overline{0}$.

{\L}ukasiewicz logic is complete with respect to deductions from finite theories for the given
semantics, i.e.: for every finite theory
$T$ and every formula $\varphi$, $T\vdash \varphi$ iff every valuation $e$ that satisfies $T$
also satisfies $\varphi$.

For each $k\in \mathbb{N}$, the finite-valued {\L}ukasiewicz logic {\L}$_k$ is the schematic extension of {\L} with the axiom schemas:\smallskip

\begin{tabular}{ll}({\L}5) $(n-1)\varphi\leftrightarrow n\varphi,\qquad \qquad$ &
({\L}6) $(k\varphi^{k-1})^n\leftrightarrow n\varphi^k$,\end{tabular}\smallskip

\noindent for each integer $k=2, \dots, n-2$ that does not divide $n-1$, and where $n\varphi$ is an
abbreviation for $\varphi\oplus\dots\oplus\varphi$ ($n$ times) and $\varphi^k$ is an abbreviation
for $\varphi\&\dots\&\varphi$, ($k$ times).
The notions of valuation and satisfiability for {\L}$_k$ are defined as above just replacing $[0, 1]$ by
\[
L_k=\left\{0, \frac{1}{k}, \dots, \frac{k-1}{k}, 1\right\}
\]as set of truth values.
Every \L$_k$ is complete (in the above sense) with respect to deductions from finite theories for the given
semantics.

It is sometimes useful to introduce
constants in addition to $\overline{0}$ that will denote values in the
domain $L_k$. Specifically, we will denote by \L$^c_k$ the
{\L}ukasiewicz logic obtained by adding constants $\overline{c}$ for
every value $c \in L_k$. We assume that valuation functions $e$
interpret such constants in the natural way: $e(\overline{c}) = c$.

A McNaughton function \cite{CiMuOt99}
is a continuous piecewise linear polynomial
functions with integer coefficients over the $n$th-cube $[0, 1]^n$.
To each {\L}ukasiewicz formula $\varphi(p_1, \dots, p_n)$ we can associate a 
McNaughton function $f_\varphi$ so that, for every valuation $e$
\[
f_\varphi(e(p_1), \dots, e(p_n))=e(\varphi(p_1, \dots, p_n)).
\]
Every {\L}-formula is then said to define a McNaughton function.
The converse is also true, i.e.
every continuous piecewise linear polynomial
function with integer coefficients over $[0, 1]^n$ is definable by a formula
in {\L}ukasiewicz logic.
In the case of finite-valued {\L}ukasiewicz logics, 
the functions defined by formulas are just the restrictions of
McNaughton functions over $(L_k)^n$.
In this sense, we can associate to every formula $\varphi(p_1, \dots, p_n)$ from {\L}$_k$
a function $f_\varphi: (L_k)^n\to L_k$. As for each \L$_k^c$, the functions defined by a formula are combinations
of restrictions of McNaughton functions and, in addition, the constant functions for each $c\in L_k$.
{The class of functions definable by \L$_k^c$-formulas exactly coincides with the class of all
functions $f: (L_k)^n\to L_k$, for every $n\geq 0$.}

The expressive power of infinite-valued \L ukasiewicz logic lies in, and is limited to, the definability of
piecewise linear polynomial functions. Expanding \L\ with the connectives $\cp, \tp$ of Product logic \cite{Ha98}, interpreted as the product of reals and as the truncated division, respectively, 
significantly augments the expressive power of the logic. 
The $\lpii$ logic \cite{EsGoMo00} is the result of this expansion,
obtained by adding the connectives $\cp, \tp, \chalf$,
whose
valuations $e$ extend the valuations for \L\ as follows:
\[
\begin{array}{r c l c c r c l c c r c l}
e(\varphi \cp \psi)& = & e(\varphi)\cdot e(\psi), & & &
e(\varphi \tp \psi)& = & \left\{
	\begin{array}{ll}
	1 & \quad e(\varphi)\leq e(\psi) \\
	\frac{e(\psi)}{e{\varphi}} & \quad \mbox{otherwise}\end{array}
	\right., & & &
e\left(\chalf\right)& = & \half.
\end{array}
\]
Notice that the presence of the constant $\chalf$ makes it possible to define constants for all
rationals in $[0, 1]$ (see \cite{EsGoMo00}).
$\lpii$'s axioms include the axioms of \L ukasiewicz and Product logics
(see \cite{Ha98}) as well as the following additional axioms, where $\Delta\varphi$ is $\neg\varphi\tp\overline{0}$:\medskip

\begin{tabular}{r l }
$(\lpi1)$ & $(\f\cp\p)\ominus(\f\cp\x)\leftrightarrow\f\cp(\p\ominus\x)$,\\
$(\lpi2)$ & $\Delta(\varphi\to \psi)\to (\varphi\tp \psi)$,\\
$(\lpi3)$ & $\Delta(\varphi\tp \psi)\to (\varphi\to \psi)$,\\
$(\lpi4)$ & $\chalf\leftrightarrow\neg\chalf$.
\end{tabular}\medskip

\noindent The deduction rules are modus ponens for $\conj$ and $\to$, and the necessitation
rule for $\Delta$, i.e.: from $\varphi$ derive $\Delta\varphi$.
$\lpii$ is complete with respect to deductions from finite theories for the given
semantics \cite{EsGoMo00}.

While \L\ is the logic of McNaughton functions, $\lpii$ is the logic of piecewise rational functions
over $[0, 1]^n$, for all $n$ (see \cite{MoPa01}). In fact, the function defined by each $\lpii$-formula with $n$ variables
corresponds to a supremum of rational fractions
\[
\frac{P(x_1, \dots, x_n)}{Q(x_1, \dots, x_n)}
\]
over $[0, 1]^n$, where $P(x_1, \dots, x_n), Q(x_1, \dots, x_n)$ are polynomials with rational coefficients.
Conversely, every piecewise rational function with over the unit cube $[0, 1]^n$
can be defined by an $\lpii$-formula.

\section{{Logics for {\L}ukasiewicz Games with Expectations}}

In this section we briefly introduce \L ukasiewicz games on {\L}$^c_k$
along with the logics $\mathsf{E}(\mathfrak{G})$ to represent
expected payoffs in classes of games. $\mathsf{E}(\mathfrak{G})$ will be the basis upon which Expectation Games are defined.

\subsection{\L ukasiewicz Games}

\begin{definition}[\cite{MaWo13}]
A {\L}ukasiewicz game $\mathcal{G}$ on {\L}$^c_k$ is a tuple 
$\mathcal{G}=\langle \sfP, \sfV, \{\sfV_i\}, \{\sfS_i\}, \{\varphi_i\}\rangle$
where:
\begin{enumerate}
\item $\sfP=\{P_1, \dots, P_n\}$ is a set of {\em players};

\item $\sfV=\{p_1, \dots, p_m\}$ is a finite set of propositional variables; 

\item For each $i\in \{1, \dots, n\}$, $\sfV_i\subseteq \sfV$ is the set of propositional variables under control of player $P_i$, so that the sets $\sfV_i$
form a partition of $\sfV$, {with} $|\sfV_i|=m_i$, and $\sum_{i=1}^nm_i=m$.

\item For each $i\in \{1, \dots, n\}$, $\sfS_i$ is the strategy set for player $P_i$ that consists of all valuations $s:\sfV_i\to L_k$
of the propositional variables in $\sfV_i$, i.e. $\sfS_i=\{s\mid s:\sfV_i\to L_k\}.$

\item For each $i\in \{1, \dots, n\}$, $\varphi_i(p_1, \dots,p_t)$ is an {\L}$^c_k$-formula, built from variables in $\sfV$, 
whose
associated function
$f_{\varphi_i}: (L_k)^t\to L_k$
corresponds to the {\em payoff function} of $P_i$, and
whose value is determined
by the valuations in $\{\sfS_1, \dots, \sfS_n\}$.
\end{enumerate}
\end{definition}

We denote by $\sfS=\sfS_1\times\cdots\times\sfS_n$ the product of the strategy spaces.
A tuple $\vec{s}=(s_1, \dots, s_n)\in \sfS$ of strategies is called a {\em strategy combination}.
With an abuse of notation, we denote by
$f_{\varphi_i}(\vec{s})$
the value of the payoff function $f_{\varphi_i}$ under the valuation corresponding to the
strategy combination $\vec{s}$.

Given a game $\mathcal{G}$, let
$\delta:\sfP\to \{1, \dots, m\}$
be a function assigning to each player $P_i$ an integer from $\{1, \dots, m\}$
that corresponds to the number of variables in $\sfV_i$: i.e.:
$\delta(P_i)=m_i.$
$\delta$
is called a {\em variable distribution function}.
Given a game $\mathcal{G}$, the {\em type} of $\mathcal{G}$ is the triple
$\langle n, m, \delta\rangle$, where $n$ is the number of players, $m$
is the number of variables in $\sfV$, and $\delta$
is the variable distribution function for $\mathcal{G}$.

\begin{definition}[Class]
Let $\mathcal{G}$ and $\mathcal{G}'$ be two {\L}ukasiewicz games $\mathcal{G}$ and $\mathcal{G}'$ on
$\mathrmL^c_k$
of type $\langle n, m, \delta\rangle$ and $\langle n, m, \delta'\rangle$,
respectively.
We say that $\mathcal{G}$ and $\mathcal{G}'$
belong to the same class $\mathfrak{G}$
if there exists a permutation $\mathfrak{j}$ of the indices $\{1, \dots, n\}$ such that,
for all $P_i$, $\delta(P_{\mathfrak{j}(i)})=\delta'(P_i).$
\end{definition}

\noindent Notice that what matters in the definition of a type is not which players are assigned certain variables,
but rather their distribution.

Let $\mathcal{G}$
be a {\L}ukasiewicz game on {\L}$^c_k$.
A {\em mixed strategy} $\pi_i$ for player $P_i$ is a probability distribution on the strategy space $\mathsf{S}_i$.
By $\pi_{-i}$, we denote the tuple of mixed strategies
$(\pi_1, \dots, \pi_{i-1}, \pi_{i+1}, \dots, \pi_n)$.
$P_{-i}$ denotes the tuple of players $(P_1, \dots, P_{i-1}, P_{i+1}, \dots, P_{n})$.
Given the mixed strategies $(\pi_1, \dots, \pi_n)$,
the {\em expected payoff} for $P_i$ of playing $\pi_i$, when $P_{-i}$ play $\pi_{-i}$, is given by
\[
exp_{\varphi_i}(\pi_i, \pi_{-i})= \underset{\vec{s}=(s_1, \dots, s_n) \in \mathsf{S}}\sum\left(\left(\underset{{j=1}}{\overset{n}\prod} \pi_j(s_j)\right)\cdot f_{\varphi_i}\left(\vec{s}\right)\right)
\]

\subsection{The Logics $\mathsf{E}(\mathfrak{G})$}

Given a class of games $\mathfrak{G}$ on $\mathrmL_k^c$, the language of $\mathsf{E}(\mathfrak{G})$ is defined as follows:
$(1)$ The set NModF of non-modal formulas corresponds to the set of $\mathrmL_k^c$-formulas built from the propositional variables $p_1,\dots, p_m$.
$(2)$ The set ModF of modal formulas is built from the atomic modal formulas
$\mathsf{E}\varphi$,
with $\varphi\in$ NModF, 
using the connectives of the $\lpii$ logic.
$\mathsf{E}\varphi$ is
meant to encode a player's expected payoff of playing a mixed strategy, given the payoff function associated to $\varphi$.
Nested modalities are not allowed.

A model $\mathbf{M}$ for $\mathsf{E}(\mathfrak{G})$ is a tuple $\langle \sfS, e, \{\pi_i\}\rangle$, such that:

\begin{enumerate}

\item $\sfS=\sfS_1\times\cdots\times\sfS_n$ is the set of all strategy combinations, i.e.
\[
\{\vec{s}=(s_1, \dots, s_n)\mid (s_1, \dots, s_n)\in \sfS_1\times\cdots\times\sfS_n\}.
\]

\item $e:({\rm NModF}\times \sfS)\to {L_k}$ is a valuation of non-modal formulas, such that, for each $\varphi\in$ NModF
$e(\varphi, \vec{s})=f_{\varphi}(\vec{s})$,
where $f_{\varphi}$ is the function associated to $\varphi$ and $\vec{s}=(s_1, \dots, s_n)$.

\item $\pi_i:\sfS_i\to [0, 1]$ is a probability distribution, for each $P_i$.

\end{enumerate}

\noindent The truth value of a formula $\Phi$ in $\mathbf{M}$ at $\vec{s}$, denoted $\|\Phi\|_{\mathbf{M},\vec{s}}$, is inductively defined as follows:

\begin{enumerate}
\item If $\Phi$ is a non-modal formula $\varphi\in$ NModF, then 
$\|\varphi\|_{\mathbf{M},\vec{s}}=e(\varphi, \vec{s})$,
\item If $\Phi$ is an atomic modal formula $\mathsf{E}\varphi$, then
$\| \mathsf{E}\varphi \|_{\mathbf{M},\vec{s}}=exp_{\varphi}(\pi_1, \dots, \pi_n).$

\item If $\Phi$ is a non-atomic modal formula, its truth value is computed by evaluating its atomic modal subformulas and then by using the truth functions associated to the $\lpii$-connectives occurring in $\Phi$.
\end{enumerate}

Since the valuation of a modal formula $\Phi$ does not depend on a specific strategy combination but only on the model
$\mathbf{M}$, we will often simply write
$\| \Phi \|_{\mathbf{M}}$ to denote the valuation of $\Phi$ in $\mathbf{M}$.

\begin{theorem}[Completeness]\label{completeness}
Let $\Gamma$ and $\Phi$ be a finite modal theory and a modal formula in $\mathsf{E}(\mathfrak{G})$.
Then,
$\Gamma \vdash_{\mathsf{E}(\mathfrak{G})} \Phi$
if and only if for every model $\mathbf{M}$ such that, for each
$\Psi\in \Gamma$, $\|\Psi\|_{\mathbf{M}}=1$, also $\|\Phi\|_{\mathbf{M}}=1$.
\end{theorem}

\section{Expectation Games}

In this section we introduce a class of games with polynomial constraints over expectations.
These games expand Lukasiewicz games by assigning to each player a formula $\Phi_i$ of $\mathsf{E}({\mathfrak{G}})$, whose interpretation
corresponds to a piecewise rational polynomial function whose variables are expected values. The formula $\Phi_i$ is meant to represent a player's goal concerning the relation between her and other players' expectations.

\begin{definition}
An Expectation Game $\mathcal{E}_\mathcal{G}$ on $\mathsf{E}({\mathfrak{G}})$ is a tuple
$\mathcal{E}_\mathcal{G}=\langle \mathcal{G}, \{\sfM_i\},\{\Phi_i\}\rangle$,
where:
\begin{enumerate}
\item $\mathcal{G}$ is a \L ukasiewicz game on $\mathrmL_k^c$,
with $\mathcal{G}\in\mathfrak{G}$,
\item for each $i\in \{1, \dots, n\}$, $\sfM_i$ is the set of all mixed strategies on $\sfS_i$ of player $P_i$,
\item for each $i\in \{1, \dots, n\}$, $\Phi_i$ is an $\mathsf{E}({\mathfrak{G}})$-formula such that every atomic modal formula occurring in $\Phi_i$
has the form $\sfE\psi$, with $\psi\in \{\varphi_1, \dots, \varphi_n\}$, i.e. the payoff formulas in $\mathcal{G}$.
\end{enumerate}
\end{definition}

A model
$\mathbf{M}=\langle \sfS, e, \{\pi_i\}\rangle$ of $\mathsf{E}(\mathfrak{G})$ for a game $\mathcal{E}_\mathcal{G}$ is called
a {\em best response model} for a player $P_i$ whenever, for all models
$\mathbf{M}'=\langle \sfS, e, \{\pi'_i\}\rangle$ with $\pi'_{-i}=\pi_{-i}$,
\[
\|\Phi_i\|_{\mathbf{M}'}\leq \|\Phi_i\|_\mathbf{M}.
\]

An expectation game $\mathcal{E}_\mathcal{G}$ on $\mathsf{E}({\mathfrak{G}})$ is said to have a {\em Nash Equilibrium},
whenever there exists a model $\mathbf{M}^*$ that is a best response model for each player $P_i$.
In that case $\mathbf{M}^*$ is called an {\em equilibrium model}.\smallskip

\noindent {\bf Example 1}. Let $\mathcal{E}_\mathcal{G}$ be any expectation game where
each $P_i$  is simply assigned the formula $\Phi_i:=\sfE\varphi_i$. This game corresponds to the
the situation where each player cares only about her own expectation and whose goal is its maximisation.
Clearly, by Nash's Theorem \cite{Na51},
every $\mathcal{E}_\mathcal{G}$ of this form admits an Equilibrium, since it offers a
formalisation of the classical case where equilibria are given by tuples of mixed strategies over valuations
in a {\L}ukasiewicz game.\medskip

\noindent {\bf Example 2}. Not every expectation game has an equilibrium. In fact, consider the following game
$\mathcal{E}_\mathcal{G}=\langle \sfP, \sfV, \{\sfV_i\}, \{\sfS_i\}, \{\varphi_i\}, \{\sfM_i\},\{\Phi_i\}\rangle$,
with $i\in \{1, 2\}$, where:\smallskip
\begin{center}
$(1)$ $\varphi_1:=p_1$ and $\varphi_2:=p_2,\qquad$ and
$\qquad(2)$ $\Phi_1:=\neg d(\sfE(p_1), \sfE(p_2))$ and $\Phi_2:= d(\sfE(p_1), \sfE(p_2))$.\footnote{ Where $\neg d(\sfE(p_1), \sfE(p_2))$ is interpreted as $1-|exp_{p_1}(\pi_1, \pi_2)-exp_{p_2}(\pi_1, \pi_2)|$
and $d(\sfE(p_1), \sfE(p_2))$ as $|exp_{p_1}(\pi_1, \pi_2)-exp_{p_2}(\pi_1, \pi_2)|$ (see \cite{GoMa13}).}
\end{center}\smallskip
\noindent The above game can be regarded as a particular version of Matching Pennies with expectations. In fact, while $P_1$
aims at matching $P_2$'s expectation, $P_2$ wants their expectations to be as far as possible. 
It is easy to see that there is no model $\mathbf{M}$ that gives
an equilibrium for $\mathcal{E}_\mathcal{G}$. Therefore:

\begin{proposition}
There exist Expectation Games on $\mathsf{E}({\mathfrak{G}})$ that do not admit a Nash Equilibrium.
\end{proposition}

\section{Complexity}

\begin{definition}
For a given game 
$\mathcal{E}_\mathcal{G}$, the \textsc{Membership} problem is the problem of determining  whether there exists an equilibrium model $\mathbf{M}$.
For a given game 
$\mathcal{E}_\mathcal{G}$ and model $\mathbf{M}$ with with rational mixed strategies
$(\pi_1, \dots, \pi_n)$, the \textsc{Non-Emptiness} problem is the problem of determining  whether
$\mathbf{M}$ belongs to the set of Nash Equilibria.
\end{definition}

Recall that the first-order theory $\mathsf{Th}(\mathbb{R})$ of real closed fields is the set of sentences in 
the language of ordered rings $\langle +, -, \cdot, 0, 1, <\rangle$ that are valid over the field of reals \cite{Ho93}.
The existence of an equilibrium in a 
game $\mathcal{E}_\mathcal{G}$ can be expressed through a first-order sentence $\xi$ of $\mathsf{Th}(\mathbb{R})$:

\begin{proposition}
For each Expectation Game $\mathcal{E}_\mathcal{G}$ there exists a first-order sentence $\xi$
of the theory $\mathsf{Th}(\mathbb{R})$ of real closed fields so that
$\mathcal{E}_\mathcal{G}$ admits a Nash Equilibrium if and only if $\xi$
holds in $\mathsf{Th}(\mathbb{R})$.
\end{proposition}

\noindent As a consequence of the above, it is easy to see that a game $\mathcal{E}_\mathcal{G}$ admits an equilibrium
if and only if there exists a quantifier-free formula in the language of ordered rings that defines
a non-empty semialgebraic set over the reals \cite{Ho93}.

We exploit the connection with $\mathsf{Th}(\mathbb{R})$ to determine the computational complexity of both 
the \textsc{Membership} and the \textsc{Non-Emptiness} problem.
In fact, given a game $\mathcal{E}_\mathcal{G}$, it can be shown that the sentence
$\xi$ can be computed from $\mathcal{E}_\mathcal{G}$ but its length is exponential
in the number of propositional variables of the payoff formulas $\phi_i$.
Deciding the validity of a sentence in $\mathsf{Th}(\mathbb{R})$
is singly exponential in the number of variables and doubly exponential in the
number of alternations of quantifier blocks \cite{Gr88}.
It can be shown that for every game the alternation of quantifiers in $\xi$
is always fixed. As a consequence, we obtain:

\begin{theorem}
Given an Expectation Game $\mathcal{E}_\mathcal{G}$ the \textsc{Non-Emptiness} problem
can be decided in {\em 2-EXPTIME}.
\end{theorem}

Deciding the validity of a sentence with only existential quantifiers in $\mathsf{Th}(\mathbb{R})$
can be solved in PSPACE \cite{Ca88}. We can show that, given a game
$\mathcal{E}_\mathcal{G}$ and model $\mathbf{M}$ with rational mixed strategies $(\pi_1, \dots, \pi_n)$,
we can compute in polynomial time an existential sentence of $\mathsf{Th}(\mathbb{R})$
whose validity is equivalent to the fact that $\mathbf{M}$ is an equilibrium model.

\begin{theorem}
Given an Expectation Game $\mathcal{E}_\mathcal{G}$ and a model $\mathbf{M}$ with rational mixed strategies
$(\pi_1, \dots, \pi_n)$, the \textsc{Membership} problem
can be decided in {\em PSPACE}.
\end{theorem}

\section{Extensions and Future Work}

This work lends itself to several extensions and generalizations.
On the one hand we plan to study the notion of correlated equilibria for Expectation Games as well as
to determine the complexity of checking their existence.
In addition, we are interested in studying games where an external agent can exert influence on the game
by imposing constraints on the payoffs and the expectations. This agent would then play the role of an enforcer
by pushing the players to make choices that agree with her dispositions.
Also, we plan to investigate games based on infinite-valued \L ukasiewicz logic \cite{CiMuOt99}
where players have infinite strategy spaces.
Finally, we intend to explore possible relations with stochastic games and whether our framework can 
be adapted to formalize those kinds of strategic interactions.

\section*{Acknowledgements}

Godo acknowledges support from the
Spanish projects EdeTRI (TIN2012-39348-C02-01) and AT (CONSOLIDER CSD 2007-0022).
Marchioni acknowledges support from the Marie Curie Intra-European Fellowship NAAMSI (FP7-PEOPLE-2011-IEF).

\bibliographystyle{eptcs}
\bibliography{generic}

\end{document}